\documentclass[10pt, superscriptaddress, bibnotes, amsmath,amssymb, aps, pre, twocolumn]{revtex4-2}
\usepackage[english]{babel}
\usepackage{graphicx}
\usepackage{dcolumn}
\usepackage{bm}
\usepackage{hyperref}
\usepackage{xcolor}

\begin{document}

\title{\textbf{Modeling multistability and hysteresis in urban congestion spreading} }%

\author{Jung-Hoon Jung}
\affiliation{Department of Physics, University of Seoul, Seoul 02504, Republic of Korea}
\author{Young-Ho Eom}
 \email{yheom@uos.ac.kr}
\affiliation{Department of Physics, University of Seoul, Seoul 02504, Republic of Korea}
\affiliation{Urban Big data and AI Institute, University of Seoul, Seoul 02504, Republic of Korea}


\begin{abstract}
Growing evidence suggests that the macroscopic functional states of urban road networks exhibit multistability and hysteresis, but microscopic mechanisms underlying these phenomena remain elusive. Here, we demonstrate that in real-world road networks, the recovery process of congested roads is not spontaneous, as assumed in existing models, but is hindered by connected congested roads, and such hindered recovery can lead to the emergence of multistability and hysteresis in urban congestion dynamics. By analyzing real-world urban traffic data, we observed that congestion propagation between individual roads is well described by a simple contagion process like an epidemic, but the recovery rate of a congested road decreases drastically by the congestion of the adjacent roads unlike an epidemic. Based on this microscopic observation, we proposed a simple model of congestion propagation and dissipation, and found that our model shows a discontinuous phase transition between macroscopic functional states of road networks when the recovery hindrance is strong enough through a mean-field approach and numerical simulations. Our findings shed light on an overlooked role of recovery processes in the collective dynamics of failures in networked systems.
\end{abstract}

\maketitle

\newcommand{\state}[2]{s^{#1}_{#2}}
\newcommand{\adjcong}{\theta}
\newcommand{\str}[2]{\adjcong^{#1}_{#2}}
\newcommand{\hinder}[0]{\xi}

Every day, urban road networks carry millions of people and the flow of traffic on them oscillates between severely congested states and well-functioning states. Understanding the nature of this recurrent transition is essential for mitigating urban traffic congestion. 

Recent empirical studies analyzing city-scale traffic data suggest that the macroscopic functional state of urban road networks exhibits multistability with associated hysteresis using the macroscopic fundamental diagram~\cite{geroliminis2008ExistenceUrbanscaleMacroscopic, geroliminis2011HysteresisPhenomenaMacroscopic},
percolation approach~\cite{zeng2020MultipleMetastableNetwork,
zeng2019SwitchCriticalPercolation,
daqing2014SpatialCorrelationAnalysis,
ambuhl2023UnderstandingCongestionPropagation, 
kwon2023GlobalEfficiencyNetwork,
li2015PercolationTransitionDynamical,
li2021PercolationComplexNetworks},
and congestion spreading patterns~\cite{duan2023SpatiotemporalDynamicsTraffic, 
jung2023EmpiricalAnalysisCongestion}.
For example, the total number of vehicles per time on a given urban road network is similar during the morning and evening rush hours~\cite{geroliminis2008ExistenceUrbanscaleMacroscopic, geroliminis2011HysteresisPhenomenaMacroscopic, saberi2013HysteresisCapacityDrop, taillanter2021EmpiricalEvidenceJamming},
but the spatial or temporal patterns of urban congestion are different depending on  whether it is in the breakdown or recovery process~\cite{geroliminis2011HysteresisPhenomenaMacroscopic, zeng2020MultipleMetastableNetwork, jung2023EmpiricalAnalysisCongestion}.
Such irreversible features show why traffic congestion control is difficult once congestion arises.

In urban road networks, due to the interconnectivity of roads, congestion on a road can propagate over time and space, ultimately creating a gridlock over the entire network. Most models that reproduce the dynamics of functional states of urban road networks are based on such congestion propagation mechanisms. 
In particular, many models inspired by epidemic spreading~\cite{saberi2020SimpleContagionProcess,
chen2022QuasicontagionProcessModeling,
ambuhl2023UnderstandingCongestionPropagation,
wang2023ContagionProcessGuided,
kozhabek2024ModelingTrafficCongestion} 
or cascading failures~\cite{wu2007CascadingFailuresWeighted,
daqing2014SpatialCorrelationAnalysis,
jia2020DynamicCascadingFailure,
cwilich2022CascadingTrafficJamming,
duan2023SpatiotemporalDynamicsTraffic}
have been suggested to understand how urban road networks can fail or how large-scale congestion arises in a dynamical perspective.

Existing models, however, leave three crucial questions about the congestion dynamics of urban road networks unanswered. (i) \emph{How does local congestion actually propagate?} Epidemic spreading models assume simple contagion, in which a contagion event can occur with a single exposure, whereas cascading failure models assume complex contagion, in which multiple exposures are necessary for a contagion event. Although these two propagating mechanisms are fundamentally different~\cite{cencetti2023DistinguishingSimpleComplex}, both have been applied to model the same phenomenon: congestion propagation. 
(ii) \emph{How does local congestion actually dissipate?} Existing models underestimated the microscopic recovery process from congestion, commonly simplified as spontaneous recovery without empirical validation. However, the interconnectivity of roads can affect not only congestion propagation, but also congestion dissipation. 
(iii) \emph{How do multistability and hysteresis phenomena arise?} Existing models focus on the formation of large-scale congestion, not on the nature of the system-wide transition between the networks' functional states (e.g. network congestion density, etc.). 
Especially, while the mechanisms of multistability and hysteresis in single roads have been extensively studied~\cite{sopasakis2006StochasticModelingSimulation, borsche2012ClassMultiphaseTraffic}, those in urban road networks remain largely unexplored.

In this study, we address these questions and propose a theory that links the microscopic and macroscopic phenomena of urban congestion dynamics with a simple model of congestion propagation and dissipation based on empirical data.
We measured the transition rates (i.e., state switching probability per unit time) between the free-flow and congested states of individual roads in real-world urban road networks and observed that congestion propagation is well described by a simple contagion process, whereas congestion dissipation (i.e., recovery from congestion) is obstructed by congestion on adjacent roads.
Inspired by this microscopic observation, we proposed a simple model of congestion dynamics, in which the recovery of a given road is hindered by congestion on neighboring roads, by modifying the susceptible-infectious-susceptible (SIS) model~\cite{pastor-satorras2015EpidemicProcessesComplex}.
We show that the hindered recovery due to neighboring congestion plays a crucial role in the emergence of multistability and hysteresis in macroscopic traffic states with an analytic approach based on a mean-field theory and confirm these results with numerical simulations of our model.

\begin{figure}
    \centering
    \includegraphics[width=\columnwidth]{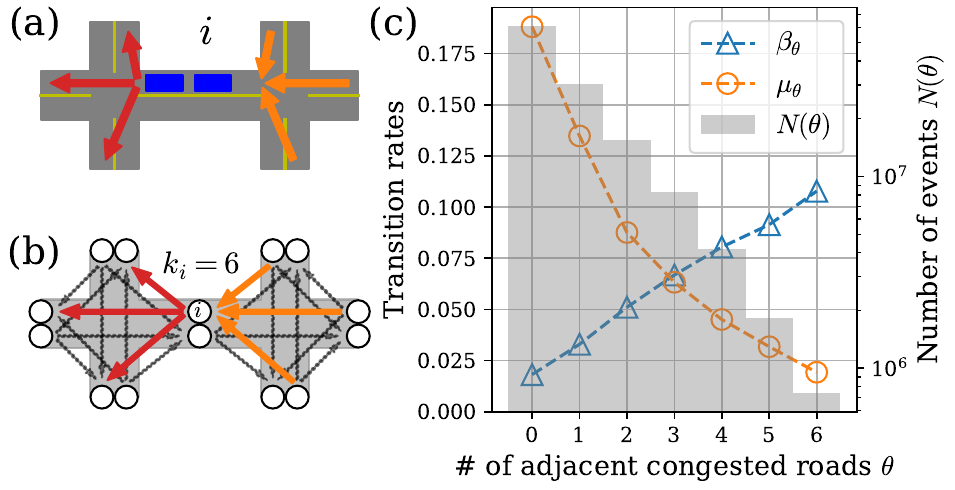}
    \caption{Measuring transition rates between traffic states of individual roads using real-world traffic velocity data.
    (a) Schematics of interactions between road segments separated by intersections in an urban road network. 
    The direction of arrows denotes the direction of vehicles from each road segments.
    The road $i$ is a upstream and downstream of roads indicated by red and orange arrows, respectively.
    (b) The representation of the road-to-road network described in (a). 
    A node (i.e., a road segment) is represented by an empty circle.
    The representative node $i$ has degree $k_i = 6$ (3 downstream and 3 upstream).
    (c) Measured average transition rates of individual roads in Seoul with respect to the number of adjacent congested road segments. 
    Blue and orange dashed lines represent the propagation and recovery rates of local congestion, respectively.  
    The grey boxes in the background represent the total number of events that a road has a certain number of adjacent congestion in the entire dataset.
    }\label{fig:trans.rate}
\end{figure}

\textit{Microscopic dynamics of congestion}--- 
We empirically investigate how the traffic state of a given road changes with the state of their connected neighboring roads. 
To do this, we consider a road-to-road network with individual roads in a given city as nodes and intersections of roads as links, known as a dual approach \cite{porta2006NetworkAnalysisUrban, jia2020DynamicCascadingFailure}.
Specifically, if it is possible to reach a road $j$ from a road $i$ by traveling through one intersection between the two roads, these roads $i$ and $j$ are connected by a link (i.e., $A_{ij}=1$) in the road-to-road network (Fig.~\ref{fig:trans.rate}(a), (b), and more details in \cite{SM}).

We assumed that congestion on adjacent roads of a given road $i$ can affect the dynamics of the road $i$.
Specifically, congestion on adjacent road of a given road can cause congestion on that road (congestion contagion). 
Conversely, if the adjacent roads of a congested road are in free-flow state, congestion on that road is likely to dissipate (recovery from congestion). 
Such dynamics of the traffic states of individual roads can be described with the contagion and recovery processes of network epidemic models~\cite{pastor-satorras2015EpidemicProcessesComplex}.

We estimated the propagation and recovery rates of real-world urban road networks using velocity data collected by GPS devices on vehicles traveling on the road networks of Seoul~\cite{jung2023EmpiricalAnalysisCongestion}, Chengdu~\cite{guo2019UrbanLinkTravel}, Seattle and New York~\cite{chen2025ForecastingSparseMovement}. 
For the estimation, we need to determine whether the traffic state $\state{i}{t}$ of a given road $i$ at time $t$ is congested $C$ (i.e., $\state i t = 1$) or free flow $F$ (i.e., $\state i t = 0$) based on these velocity data. 
We defined congestion on a single road as an abnormally low velocity state that is outside the natural range of velocity variation on that road, 
and the details of how to determine road congestion can be found in ~\cite{jung2023EmpiricalAnalysisCongestion} and ~\cite{SM}.
The congestion propagation rate $\beta_\adjcong$ and the recovery rate $\mu_\adjcong$ for a single road with the number of adjacent congested roads $\adjcong$ can be calculated as below

\begin{align}\label{eqn:estimation}
    \beta_\adjcong = P(C|F;\adjcong) =
    \dfrac{\sum_{i,t} N(\state i t = 0, \state i {t+\Delta t}= 1; \str{i}{t} =\adjcong)}{\sum_{i,t} N(\state i t = 0; \str{i}{t} = \adjcong)}, \\
    \mu_\adjcong = P(F|C;\adjcong) = \dfrac{\sum_{i,t} N(\state i t= 1, \state i {t+\Delta t} = 0; \str{i}{t} = \adjcong)}{\sum_{i,t} N(\state i t = 1;\str{i}{t} = \adjcong)},
\end{align}
where $\str{i}{t} = \sum_j A_{ij}s^j_t$, with $A$ as the adjacency matrix of the underlying road-to-road network, and $N(\cdot)$ denotes the number of events satisfying the given conditions.
These transition rates tell us about the probability per unit time that the traffic states of individual roads change with the number of their congested adjacent roads.

Figure~\ref{fig:trans.rate} (c) shows the actual transition rates between the traffic states $s^i_t$ in the Seoul road-to-road network as an example (results from other cities also show qualitatively similar tendencies~\cite{SM}). 
The linear increase in the congestion propagation rate $\beta_{\adjcong}$ to the number of adjacent congested roads $\adjcong$ suggests that local congestion spreads in a manner similar to a simple contagion process, i.e., congestion propagation independently occurs by pairwise unidirectional interactions from a congested road to a free-flow road.
It also explains why congestion propagation in real-world road networks is well described and predicted by epidemic-based models~\cite{saberi2020SimpleContagionProcess,chen2022QuasicontagionProcessModeling,kozhabek2024ModelingTrafficCongestion}.
However, recovery rates $\mu_{\adjcong}$ decrease rapidly with the number of congested adjacent roads $\adjcong$.
These observations empirically show that the recovery of individual roads from congestion is not a spontaneous process, as assumed in previous studies, but a more complex process that requires the redistribution of traffic load to its surroundings.
Then, what role do the observed propagation and recovery processes at the road level play in the dynamics of city-scale traffic congestion?

\textit{Model.}---
To answer this question, we developed a modified Susceptible-Infectious-Susceptible (SIS) model with a simple contagion yet a complex recovery process as we observed in real-world road networks, in which susceptible and infectious states of nodes represent the free-flow and congested state of roads, respectively.  
More specifically, within a time interval $\Delta t$, the congestion propagation is described by the propagation rate $\beta_\adjcong = \beta_0 + \beta \adjcong$ depending on the number of congested neighbors $\adjcong$, 
which consists of the spontaneous congestion rate $\beta_0$ and the propagation strength $\beta$. 
On the other hand, we choose the exponential form of recovery rate $\mu_\theta = \mu_0 \hinder^\theta (\hinder < 1)$ from the base recovery rate $\mu_0$ with the recovery reduction ratio $\xi$ based on the value of $\adjcong$, describing the hindrance of adjacent congestion.
In this model, we focus on the network congestion density $z$ as the order parameter, which characterizes the system’s dynamical states.

The resulting model for urban congestion dynamics is similar to the multi-process contagion models investigated in synthetic networks~\cite{majdandzic2014SpontaneousRecoveryDynamical, bottcher2017FailureRecoveryDynamical,bottcher2017CriticalBehaviorsContagion},
in that the contagion and recovery rates of node states can vary across the types of node failures.
In these models, even with low internal (i.e. spontaneous) failure density, external (i.e. propagated) failures by neighboring failures form a robust cluster if such external failures occur frequently enough, and thus the system exhibits hysteresis and discontinuous phase transitions in terms of the global failure density. 
In our model, however, the propagation and recovery rates depend on the number of adjacent congested neighbors, whereas recovery rates are determined by the type of failures in previous studies~\cite{majdandzic2014SpontaneousRecoveryDynamical, bottcher2017FailureRecoveryDynamical,bottcher2017CriticalBehaviorsContagion}. 
Despite these differences, one can expect hysteresis and discontinuous phase transitions in our model because the expected recovery time $(=(1-\mu_\theta)/\mu_\theta^2)$ can vary greatly depending on the state of the surrounding roads.
Therefore, we have performed mean-field calculations to investigate the role of recovery hindrance in the macroscopic order parameter $z$, particularly in terms of hysteresis and multistability.

\begin{figure}[]
    \centering
    \includegraphics[width=\columnwidth]{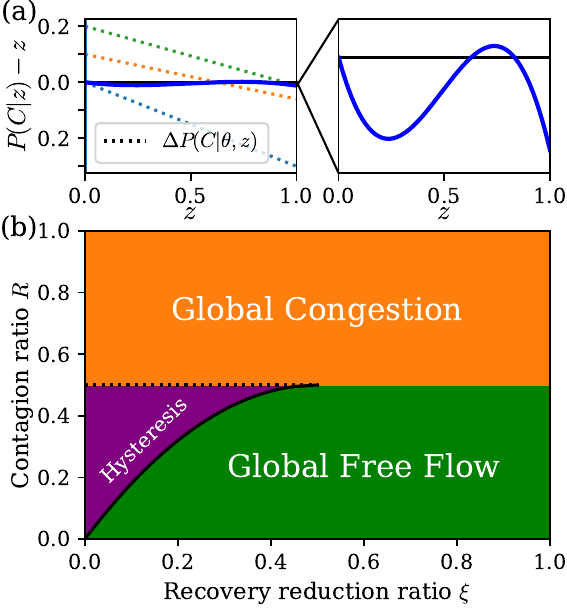}
    \caption{Representing a mean-field equation and phase diagram of our model. 
    (a) Representation of the equation of a deviation of the congestion probability $\Delta z \,(=P(C|z)-z)$ with a given congestion probability $z$.
    Each dotted line denotes the conditional deviation of congestion probability based on the number of adjacent congested roads $\adjcong$ (Blue: 0, Orange: 1, Green: 2). 
    The Blue solid line represents a total deviation of congestion probability, which is the convolution of conditional deviations and the distribution function of $\adjcong$. 
    Right panel is zooming out for the blue solid line around $\Delta z = 0$ in left panel.
    (b) Phase diagram of the mean-field analysis for $n=2$ with $\beta_0 = 0$.
    Green and orange area represent global free-flow and congestion states regardless of the initial state, respectively. 
    The purple area denotes the hysteresis region which means the global congestion density depends on the initial state.
    }
    \label{fig:mf}
\end{figure}

\textit{Mean-field approach}---
In order to examine macroscopic features of our model, we analyzed our model with a homogeneous mean-field assumption that each node (i.e., a road segment) is in the congested state $C$ independently with a certain probability $z\equiv P(C)$.
If we consider a tree-like lattice in which all nodes have the same number of adjacent roads $n$, the distribution $f_n(\adjcong|z)$ of the number of adjacent congested roads turns out to be the binomial distribution as below,
\begin{equation}\label{eqn:MF}
    f_n(\adjcong|z) = \left(
    \begin{matrix}
        n\\
        \adjcong
    \end{matrix}
    \right)
    z^\adjcong(1-z)^{n-\adjcong}.
\end{equation}
The probability $P(C|\theta, z)$ that a certain node will be in the congested state $C$ in the next time step, given the number of adjacent congested roads $\theta$ and the congestion probability $z$, can be calculated as follows,
\begin{equation}\label{eq:micro}
    P(C|\adjcong,z) = \beta_\adjcong P(F) + (1-\mu_\adjcong )P(C).
\end{equation}
The first term of eq.~(\ref{eq:micro}) represents the probability that the free-flow nodes will become congested, while the second term represents the probability that the congested nodes will not recover.

Finally, we can obtain the self-consistent equation for the congestion probability $z'\equiv P(C|z)$ in the next time step
with a given previous congestion probability $z$ can be written as below,
\begin{equation}\label{eqn:sc_eq}
    z' = P(C|z) = \sum_{\adjcong=0}^n P(C|\adjcong,z)f_n(\adjcong|z),
\end{equation}
in which a fixed point is at $z'=z$.

For the case of $n=2$, the resulting road network represents the linear road network.
To compare the original SIS model, we assume that $\beta_0 \simeq 0$.
Substituting the exponential form $\mu_0\hinder^\adjcong$ for $\mu_\adjcong$, 
one can rewrite eq.~(\ref{eqn:sc_eq}) as follows (c.f. Fig.~\ref{fig:mf}(a)),
\newcommand{\barm}{\bar \hinder}
\begin{equation}\label{eqn:n2}
    \Delta z\equiv z' - z = \mu_0 z( -\barm^2 z^2 + 2(\barm - R) z - (1 - 2R)),
\end{equation}
where $\barm = 1-\hinder$ which means the decay strength of the recovery rate, 
and the normalized contagion rate $R=\beta/\mu_0$.
One can see that the sign of $1-2R$ decides the stability of the trivial fixed point at $z=0$.
This result is consistent with the epidemic threshold in the original SIS model (c.f. $2R=2\beta/\mu_0 \simeq R_0 \equiv \langle k\rangle\beta/\mu$).

If the trivial fixed point is stable ($2R<1$), 
eq.~(\ref{eqn:n2}) has non-zero fixed points only if the determinant $D\geq0$, 
which can be written as,
\begin{equation}
    D \equiv 2\barm^2 - 2\barm + R \geq 0.
\end{equation}
This inequality gives the critical point of $\hinder$ as below,
\begin{equation}
    \hinder^*_\pm =1-\barm^*_\pm = \frac{1\mp \sqrt{1-2R}}{2}.
\end{equation}
Because the value of $\hinder^*_-$ gives the unphysical fixed point $z^* <0$,
eq.~(\ref{eqn:n2}) has the non-zero stable fixed point only when $\hinder$ satisfies below condition,
\begin{equation}
    \hinder<\frac{1-\sqrt{1-2R}}{2}.
\end{equation}
At this point, two non-trivial fixed points (one stable and one unstable) eventually emerge, indicating that the model exhibits a discontinuous phase transition and also hysteresis.
Furthermore, for general cases of $n\ge3$, 
one can show that the stability of the trivial fixed point ($z=0$) always depends on the sign of $1-nR$, and 
the non-trivial stable fixed point near $z\approx 1$ always emerges when the hindrance is strong enough ($\xi \ll 1$) because $\Delta z|_{z=1} = \hinder^n \rightarrow 0$. 
Through the mean-field analysis, we found that there is the hysteresis region in the parameter space of our model (See Fig.~\ref{fig:mf}(b)).
Although our homogeneous mean-field approach provides a tractable understanding of the macroscopic features of our model, it may fail near the critical point where local clusters emerge.

\textit{Numerical simulations}---
To confirm the mean-field results in a more realistic situation,
we have simulated our model on a synthetic bi-directional road network with the shape of the $L\times L$ square lattice.
These lattices mimic road networks found in urban areas, where each intersection has four incoming and four outgoing roads. 
Following the road-to-road transformation (See Fig.~\ref{fig:trans.rate}(b)), each node (i.e., road segment) therefore has six edges ($n =6$).
Our simulation is performed on the square lattice with $L=15$,
so that the total number of road segments is $N=900\,\,(=15\times15\times4)$.
According to the mean-field approach, we choose global recovery rate $\mu_0$ as 0.5 without loss of generality and set the spontaneous contagion rate $\beta_0 \ll 1$ as $10^{-6}$.

\begin{figure}[]
    \centering
    \includegraphics[width=0.98\columnwidth]{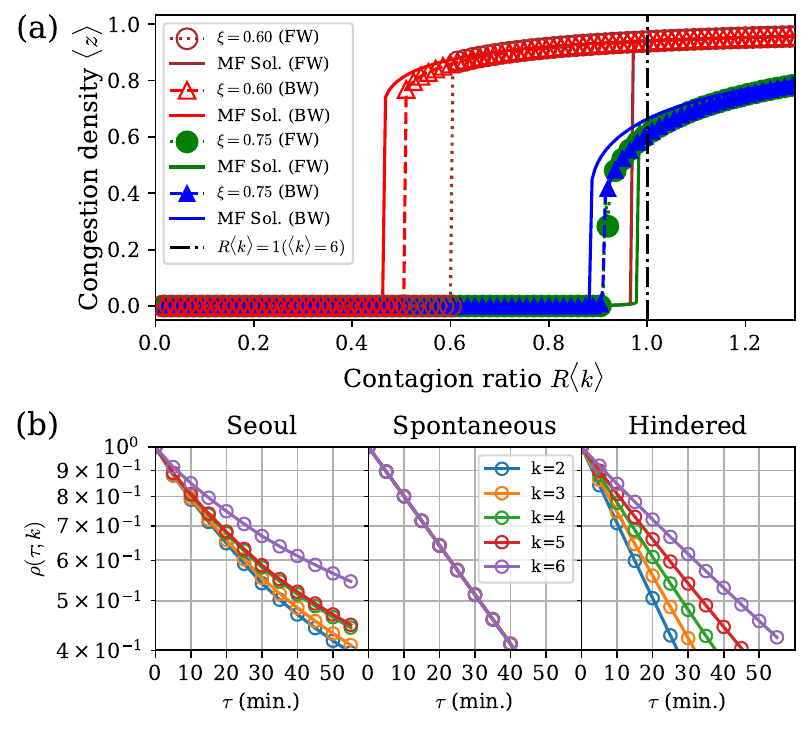}
    \caption{Numerical results of our modified SIS model.
    (a) Numerical simulation results on a square grid. 
    We have calculated the time average of the global congestion density $z$, varying the contagion ratio $R=\beta/\mu_0$ from 0 to 0.5 (forward, dotted line and circle markers) or from 0.5 to 0 (backward, dashed line and triangle markers) with $10^5$ relaxation and simulation steps.
    Each color represents a different recovery reduction ratio $\xi$ and the process of varying $R$
    (the other parameters $\beta_0$ and $\mu_0$ are set to $10^{-6}$ and 0.5, respectively).
    The colored solid lines represent the corresponding mean-field solutions of $n=3$.
    The black dash-dotted line denotes the epidemic threshold of the original SIS model which can be read as $\beta n/\mu_0  = 1$.
    (b) Autocorrelation function of empirical data of Seoul (left) and numerical simulations which are the spontaneous recovery model (middle) and the hindered recovery model (right). 
    Each color represents the degree of averaged nodes.
    }
    \label{fig:numerical}
\end{figure}

Figure~\ref{fig:numerical}(a) shows the time averaged global congestion density $\langle z \rangle$ regarded as the order parameter of our model.
The different transition points of forward and backward processes denote that hysteresis is observed in that region.
While the transition point of the backward process (dashed line) is well expected by the corresponding mean-field solution (solid line), the transition point of the forward process (dotted line) is not.
This difference between the mean-field approach and numerical simulations in the forward process seems to be the result of overestimating the stability of low-density fixed point by the mean-field approach, 
because the mean-field approach ignores the correlation of congestion, and also its local fluctuations.
These results suggest that hysteresis shown in real-world road networks might be a result of hindered recovery by adjacent congested roads, indicating that the microscopic gridlock of local congestion is more persistent than we expect, and can induce the macroscopic urban gridlock.

To examine the persistence of local congestion, observed in real-world urban networks~\cite{jung2023EmpiricalAnalysisCongestion}, we measured the $\tau$-consecutive time autocorrelation function of congestion for each node and averaged it with respect to degree, which can be calculated as below,
\begin{equation}
    \rho^i(\tau) = \dfrac {\sum_t (\prod_{\Delta t = 0}^\tau s^i_{t+\Delta t})} {\sum_t s^i_t}, \quad \rho(\tau;k) =\langle \rho^i(\tau)\rangle_{k_i = k}.
\end{equation}
For comparison, we estimated the parameters in our model and additionally the spontaneous recovery rate $\mu_s$ from the empirical data, and performed corresponding numerical simulations on the real-world road networks of Seoul and Chengdu.
Figure~\ref{fig:numerical}(b) shows the autocorrelation function of empirical data, spontaneous recovery model, and our model.
Because both numerical simulations are memory-less process, their autocorrelation functions decay exponentially.
In contrast, the autocorrelation function of empirical data exhibits much longer tails, indicating that congestion in real data contains temporal correlations.
Moreover, the empirical autocorrelation function also exhibits degree dependence, consistent with the behavior observed in our model, which is not shown in the result of the spontaneous recovery model (For the case of Chengdu, see \cite{SM}).
These results demonstrate that, despite its simplicity, our model is able to reproduce the characteristic patterns observed in empirical data.

\textit{Discussion}---
In summary, we empirically analyzed the propagation and dissipation of local congestion using real-world traffic data, and observed that the recovery mechanism of individual roads is not spontaneous but depending on the states of their neighboring roads, which previous studies have usually either overlooked or simplified as spontaneous. 
Based on this observation in the dynamics of local congestion, 
we suggested a simple model of congestion propagation and dissipation, which is the modified SIS model with nonlinearly decreasing recovery rate as a function of flow states of neighboring roads.
Our simple model provided qualitative prediction and explanation of the macroscopic functional phases, multistability, and hysteresis of urban road networks as a result of the interplay between the propagation and recovery rates.
Such hindered recovery is not limited to traffic congestion but is also related to other collective failures in networks, especially those involving load sharing for local recovery (e.g., power system~\cite{albert2004StructuralVulnerabilityNorth, dobson2007ComplexSystemsAnalysis,  wang2009ModelCascadingFailures, nesti2020EmergenceScaleFreeBlackout}, internet traffic~\cite{fu2021ModelingAnalyzingCascading, yin2023AnalysisCascadingFailures} and logistics network~\cite{yang2020ControllabilityRobustnessCascading, chen2022ResilienceLogisticsNetwork}), which is commonly observed in social infrastructure. 
Therefore, we hope that our theoretical framework with a complex recovery process can provide a new tool to better understand the emergent vulnerability and resilience of these networked systems.

Complex contagion has been highlighted in recent years for its ability to explain social phenomena~\cite{centola2007ComplexContagionsWeakness,  karsai2014ComplexContagionProcess,monsted2017EvidenceComplexContagion, sprague2017EvidenceComplexContagion, iacopini2019SimplicialModelsSocial}.
In particular, existing studies have focused on distinguishing between the propagation processes of simple and complex contagion~\cite{czaplicka2016CompetitionSimpleComplex,
min2018CompetingContagionProcesses,
cencetti2023DistinguishingSimpleComplex, 
horsevad2022TransitionSimpleComplex}. 
We have shown that complex recovery processes in urban congestion dynamics play an important role in the hysteresis and multistability of the system, 
similar to the role of complex contagion in other studies.
A comprehensive understanding of network dynamics can be obtained when considering not only activation but also deactivation process.

\begin{acknowledgments}
This work was supported by the Basic Study and Interdisciplinary R\&D Foundation Fund of the University of Seoul (2022). The authors thank Jae Dong Noh for fruitful discussions and comments on this work.
\end{acknowledgments}

\bibliography{csns20250}

\end{document}


\author{Jung-Hoon Jung}
\affiliation{
 Department of Physics, University of Seoul, Seoul 02504, Republic of Korea,
}
\author{Young-Ho Eom}
\affiliation{
 Department of Physics, University of Seoul, Seoul 02504, Republic of Korea,
}
\affiliation{Urban Big data and AI Institute, University of Seoul, Seoul 02504, Republic of Korea}
\title{Supplemental materials for ``Modeling multistability and hysteresis in urban congestion spreading"}
\maketitle
\newcommand{\jhedit}[1] {\textcolor{red}{#1}}

\section{Data Description}
To examine the dynamics of city-wide traffic congestion, we prepared the urban road velocity datasets of 4 cities, Seoul~\cite{jung2023EmpiricalAnalysisCongestion}, Chengdu~\cite{guo2019UrbanLinkTravel} (figshare: \url{https://doi.org/10.6084/m9.figshare.7140209.v4}), Seattle and New York~\cite{chen2025ForecastingSparseMovement} (github: \url{https://github.com/xinychen/tracebase/tree/main/datasets}). 
Each dataset has different spatial and temporal resolution as below. 

\setlength{\tabcolsep}{10pt}
\begin{table}[htbp]
\begin{tabular}{ 
|p{2cm}||p{2cm}|c|p{3cm}|  }

 \hline
 \multicolumn{4}{|c|}{City List} \\
 \hline
 City or State & \# of Roads & Periods & Resolution \\
 \hline
 Seoul & 4711& 2019.12.01 -- 2020.02.28& 5 min.\\ 
 \hline
 Chengdu & 5943 & 2015.06.01 -- 2015.07.15 & 2 min. \\
 \hline
 Seattle & 63351 & 2019.01.01 -- 2019.03.11 & 1 hour \\
 (filtered) & 13430 & &\\
 \hline
 New York & 98057 & 2019.01.01 -- 2019.03.31 & 1 hour\\
 (filtered) & 45928 &&\\
 \hline
\end{tabular}
 \caption{Information about datasets. \label{tab:cities}}
\end{table}

In case of Seoul and Chengdu, datasets have high temporal resolutions estimated by GPS trajectories, but contain noisy fluctuations of velocity sequences.
To reduce the noisy temporal fluctuations, we calculated the moving average for each road with a 30-minute time window.

In case of Seattle and New York (originally from Uber Movement dataset which is not available for now), datasets are hourly aggregated and have broad spatial areas which cover the whole state, but hence contain a lot of missing data points.
So, we filtered out some roads with the number of data points less than one week, and excluded missing data from the calculation of local congestion.
Because of the lack of data points, the results might be far away from the exact value of the data fully filled, however, it can be used to understand the qualitative tendency of urban congestion propagation and dissipation.

\section{Urban road-to-road network construction}
In the main text, we investigated the transition rates between traffic states of \emph{individual road segments} based on the states of adjacent road segments.
However, \emph{the usual representation} of road networks does not directly describe such adjacent relations between road segments because the conventional nodes of road networks are intersections of roads, not road segments themselves. 
To obtain the direct relation between individual road segments, we transformed each urban road networks into \emph{road-to-road} networks as a form of the dual network, 
in which road segments are identified as nodes and edges exist when a vehicle can drive from one to another road. 
Specifically, road segments which can consist of a single lane or multi lanes are identified as a single node with sharing same outgoing and incoming intersections.
If a road $i$ share an outgoing intersection with another road $j$ as its incoming intersection, then those two roads are connected by an edge from $i$ to $j$ ($A_{ij} = A_{ji} =1$).
Finally, we ignored the connection between opposite roads, so called U-turn roads, 
which is a minor part of the whole interactions between roads (e.g., U-turn is generally prohibited in highway). 
Then, the resulting network turns out a road-to-road network consist of roads as nodes and connectivity between roads as edges.

\section{Determination of traffic states based on velocity sequences}
To analyze the propagation and dissipation of local congestion in the road-to-road networks, one needs to identify whether a given road is congested or not based on the velocity of traffic flow on it.
However, because each road has a different speed limit and profile, the normalization plays an important role in the determination of local congestion,
when one identifies congestion as a less speed than a certain global threshold value.
If congestion on a road can be defined as an \emph{abnormal} slowdown beyond the natural range of velocity fluctuations in traffic flow on that road,
the distribution of a velocity sequence of each road would not follow normal statistics (e.g. bimodality, skewness or kurtosis, etc.).

Hence, in our previous work~\cite{jung2023EmpiricalAnalysisCongestion}, we have supposed that the effective z-score normalization which is the z-score normalization using quantile values rather than statistical values.
This normalization method assumed that the distribution of velocities mostly follows log-normal distribution function, 
but congestion eventually records extremely low speed values which can produce the deviation of statistical values such as the mean $\mu$ or the standard deviation $\sigma$.
So, the median and the 95th quantile value of the velocity distribution of road segments are regarded as the mean and the effective maximal value ($\mu+2\sigma$), respectively, of the typical distribution, which is the distribution without congestion.

\begin{figure}[]
    \centering
    \includegraphics[width=0.95\columnwidth]{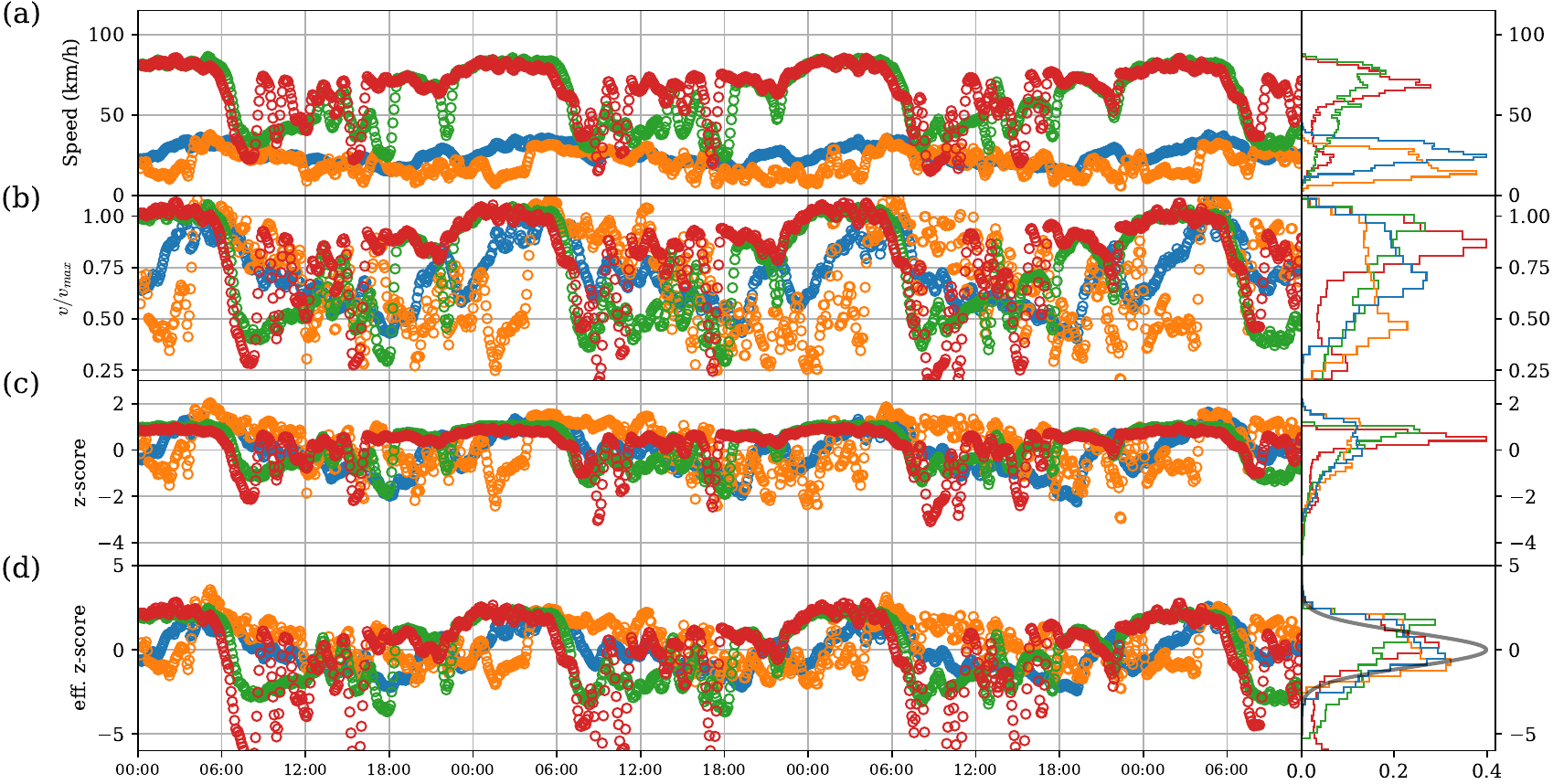}
    \caption{
    Velocity sequences and its normalizations of Seoul road segments.
    Right panel shows histograms of sequences for each colors.
    Each rows denote different standardizations of the original datasets,
    (a) original dataset, (b) the relative velocity, (c) the z-score and (d) the effective z-score, respectively.
    The gray solid line on the right panel of (d) represents the normal distribution for the comparison.
    }
    \label{fig:su_spd}
\end{figure}

\begin{figure}[]
    \centering
    \includegraphics[width=0.95\columnwidth]{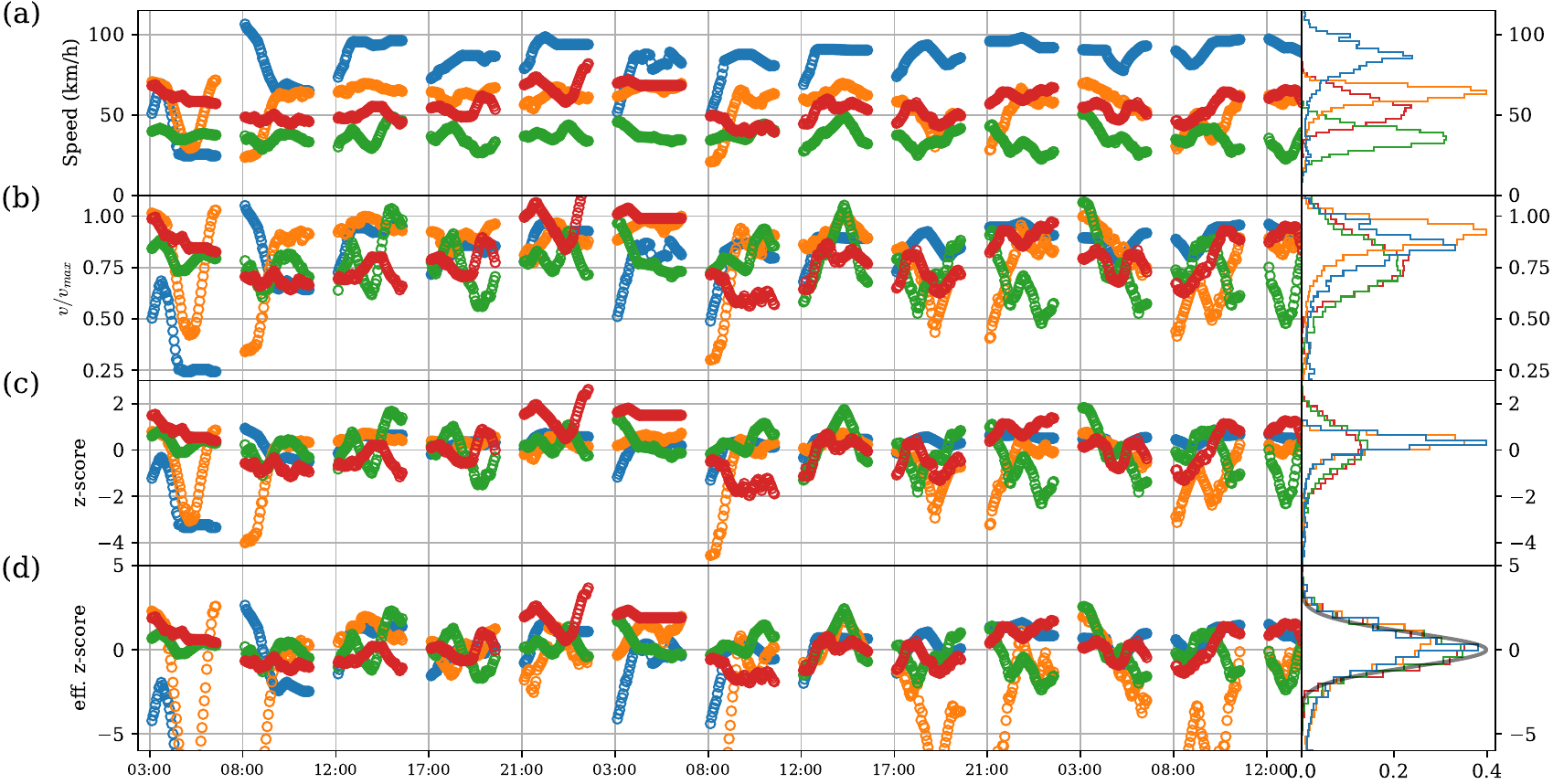}
    \caption{
    Velocity sequences and its normalizations of Chengdu road segments.
    }
    \label{fig:ch_spd}
\end{figure}

Figs.~\ref{fig:su_spd} and ~\ref{fig:ch_spd} show the velocity sequences and several types of normalization method for the comparison. 
Note that the time period of Chengdu data is 2 hours per each period.
Fig.~\ref{fig:su_spd}(b) and \ref{fig:ch_spd}(b) represent the widely used simple normalization method, the relative velocity, 
which is the velocity divided by the maximal velocity identified as the 95th quantile value.
This method provides a good quality of the normalized values, 
but generally overestimates the performance of the roads with high-speed limits.  
The z-score normalization which is the deviation from the mean divided by the standard deviation 
shows that the abnormal slowdown can affect the basic statistics of the velocity distribution (i.e., mean, standard deviation, etc.), 
which means that it is not proper standardization to determine local congestion.
In contrast, the effective z-score, which is shown in Fig.~\ref{fig:su_spd}(d) and \ref{fig:ch_spd}(d), 
shows the best quality of the empirical standardization according to the comparison with the black solid line which represents the normal distribution. 
So, we have converted all datasets we obtained into the effective z-score and determined the emergence of local congestion by a certain global threshold $h$.
We choose $h$ as $-0.5$ for the microscopic transition rate calculation in the main text.


\section{Empirical microscopic transition rates in urban road networks}
Through the datasets of local congestion on each road with a certain threshold $h$,
we calculate the microscopic transition rates between the congested state and the free-flow state by the following procedures.
First, we calculate the number of congested neighbors of each road through the adjacent matrix of the road-to-road network described in the main text and above ($\theta_i = \sum_j A_{ij}s_j$).
For each consecutive time $t$ and $t+\Delta t$ in the dataset, 
we classified and gathered the number of events of road traffic states based on the number of congested roads ($C\rightarrow C, C\rightarrow F, F\rightarrow C ~~\textrm{and}~~ F\rightarrow F$). 
We measured the microscopic transition rates, which are the propagation rate and the recovery rate based on the ratio of the number of corresponding events.

\begin{figure}[]
    \centering
    \includegraphics[width=0.95\columnwidth]{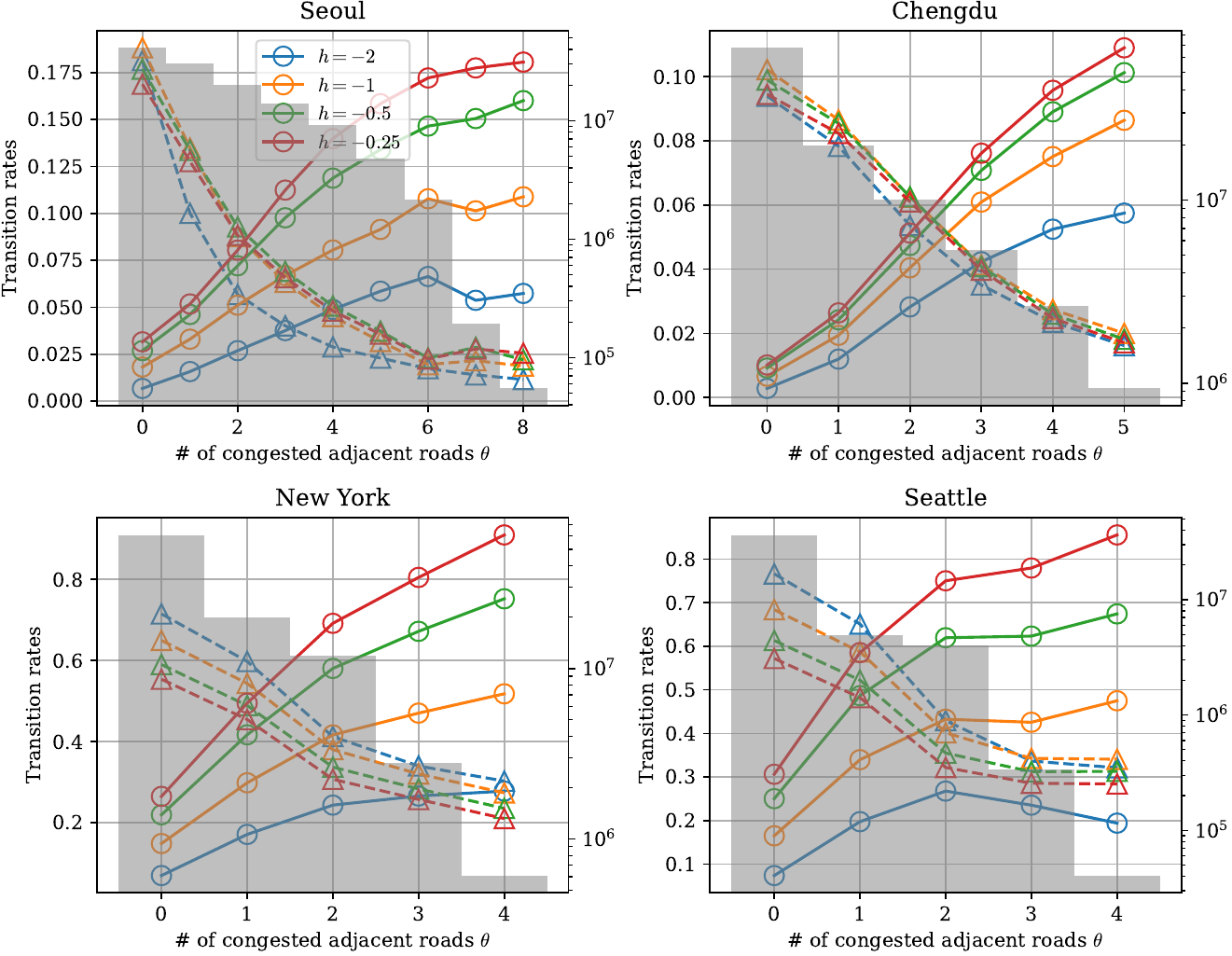}
    \caption{
    Microscopic transition rates between traffic states of individual roads for different cities.
    Each solid lines and dashed lines are representing the contagion rate and the recovery rates of local congestion based on the number of adjacent congested roads.
    The gray histogram on background shows the number of events that a certain road have neighboring congested roads $\theta$ when the global congestion threshold $h$ is given as -0.5 (c.f. the result of green in transition rates).
    }
    \label{fig:micro}
\end{figure}

Fig.~\ref{fig:micro} shows the microscopic transition rates for each city based on various congestion thresholds $h$ represented as each color.
Even though New York and Seattle have much larger number of roads in datasets, 
the actual number of events is much less than Seoul and Chengdu due to the longer time scale and a lot of missing data points, and that's the reason why they show somehow more noisy results. 
Even in some noisy results, the difference shown in the recovery rates is crystal-clear, which means congestion on roads can interrupt the recovery process of upstream congested roads. 
In this context, the congestion threshold does not induce qualitative differences among the datasets; it only affects the transition rates quantitatively.

\begin{figure}[h]
    \centering
    \includegraphics[width=0.72\columnwidth]{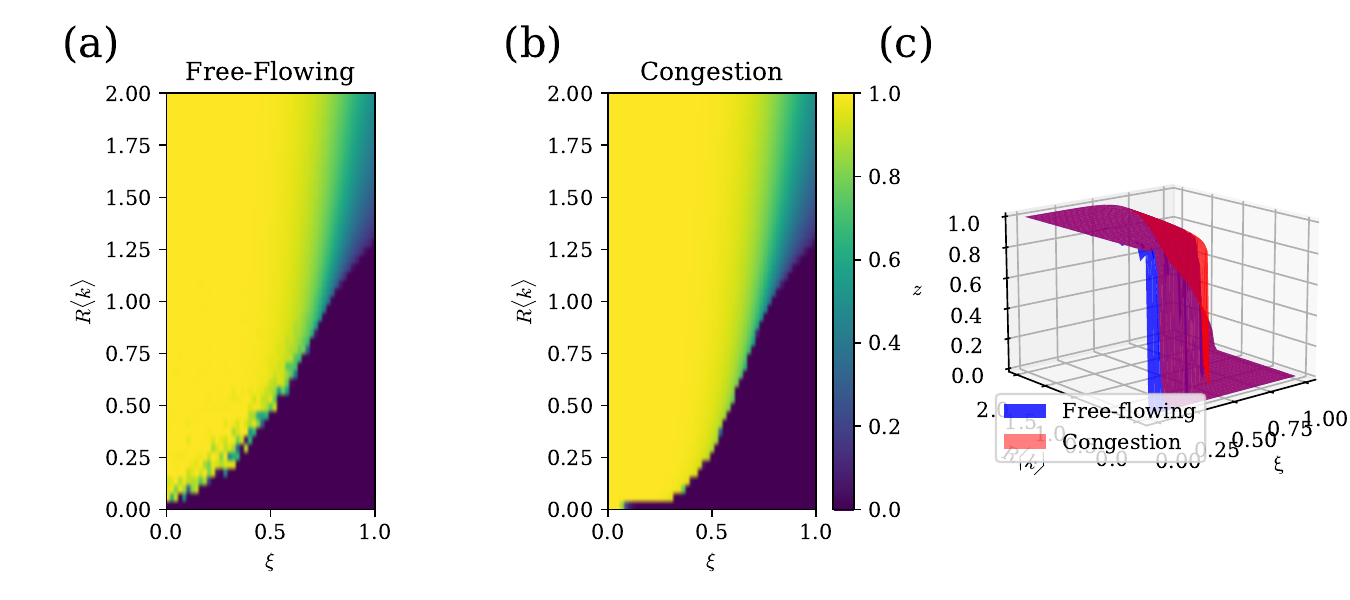}
    \caption{
    Numerical simulation results of the square grid network. 
    We simulated each parameters with different initial conditions, (a) the global free-flow state and (b) the global congested state.
    (c) Representation on 3D space of numerical results.
    Blue and Red surface represents each initial condition, the global free-flow state and the global congested state, respectively.
    }
    \label{fig:sqaure}
\end{figure}

\begin{figure}[]
    \centering
    \includegraphics[width=0.72\columnwidth]{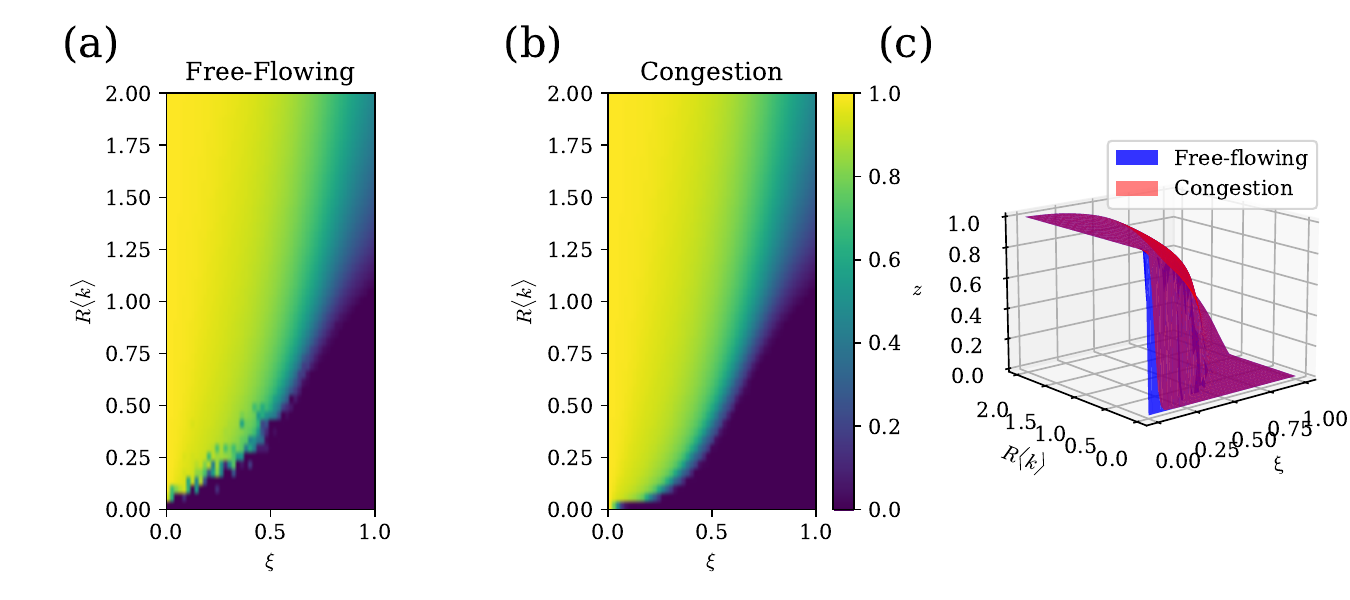}
    \caption{
    Numerical simulation results of Seoul road network. 
    We simulated each parameters with different initial conditions, (a) the global free-flow state and (b) the global congested state.
    (c) Representation on 3D space of numerical results.
    Blue and Red surface represents each initial condition, the global free-flow state and the global congested state, respectively.
    }
    \label{fig:Seoul}
\end{figure}

\begin{figure}[]
    \centering
    \includegraphics[width=0.72\columnwidth]{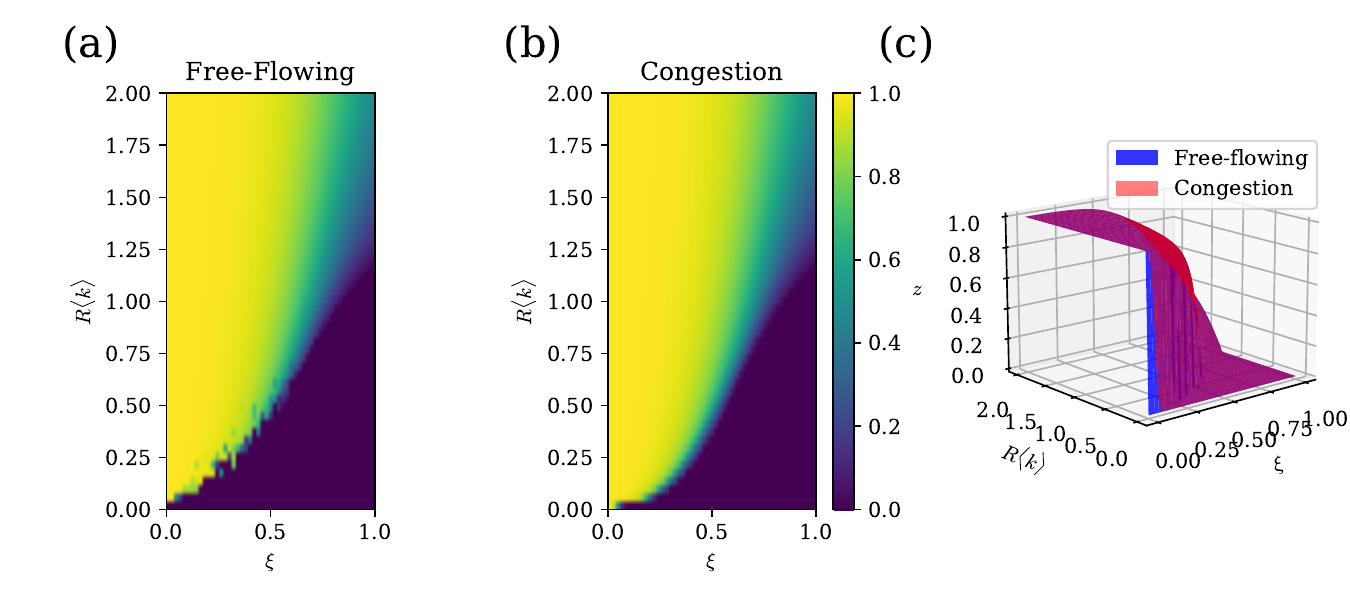}
    \caption{
    Numerical simulation results of Chengdu road network. 
    We simulated each parameters with different initial conditions, (a) the global free-flow state and (b) the global congested state.
    (c) Representation on 3D space of numerical results.
    Blue and Red surface represents each initial condition, the global free-flow state and the global congested state, respectively.
    }
    \label{fig:Chengdu}
\end{figure}

\section{Numerical simulation results}
In the main text, we showed the phase diagram of the mean-field approach of the system. 
To validate the mean-field approach, we tested our model on the square grid road network $(N = 15\times15\times4)$ and empirical road networks, Seoul and Chengdu. 
Note that the road-to-road network of a square grid road network is \emph{not} the square lattice.

We setup the parameter $\beta_0$ and $\mu_0$  as  $10^{-6}$ and $0.5$, respectively.
Each simulation has calculated $5\times 10^5$ simulation time step after $10^4$ relaxation step. 
We examined how the order parameter changed with each parameter and with different initial conditions, in order to observe the existence of hysteresis.

\begin{figure}[]
    \centering
    \includegraphics[width=0.95\columnwidth]{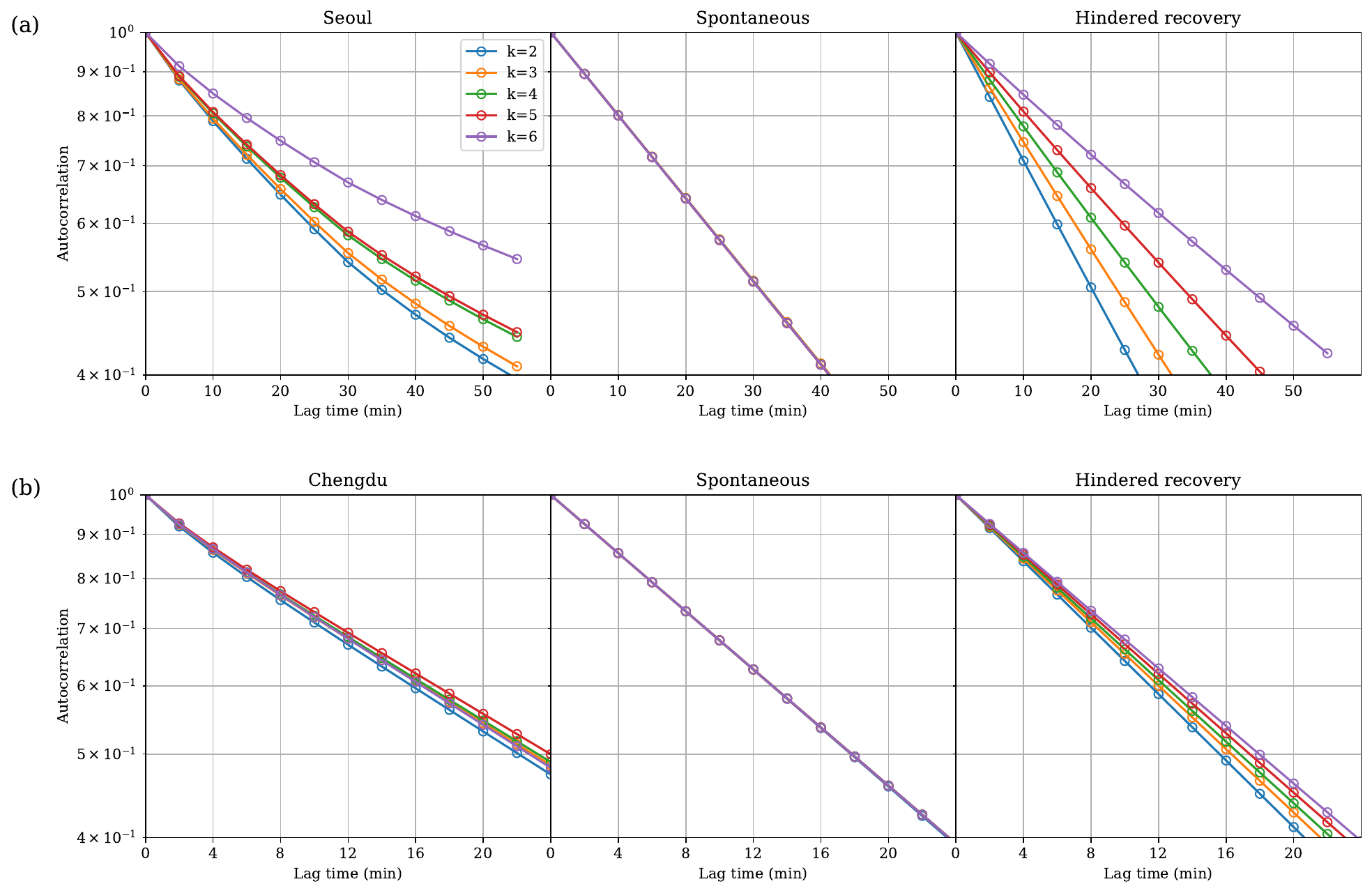}
    \caption{
    The results of $\tau$-consecutive time autocorrelation functions of (a) Seoul and (b) Chengdu.
    Each columns represents the type of congestion sequences which are empirical dataset (left), numerical simulation of the spontaneous recovery model (middle) and our hindered recovery model (right).
    Each color represents the degree of averaged nodes for each autocorrelation functions.
    }
    \label{fig:autocorr}
\end{figure}

\section{Parameter estimation for calculating the autocorrelation function}
To simulate the spontaneous recovery model and the hindered recovery model corresponding to empirical data, 
we first calculate the number of events from the empirical data and measure $\beta_\theta$ and $\mu_\theta$.  
For the hindered recovery model, two parameters, $\beta$ and $\xi$, are required. 
We estimate each parameter by fitting $\beta_\theta - \beta_0$ to $\beta \theta$ for $\beta$, 
and $\log(\mu_\theta/\mu_0)$ to $-\xi \theta$ for $\xi$.  
For the spontaneous recovery model, we additionally obtain the spontaneous recovery rate $\mu_s$ without degree dependence by calculating the fraction of recovery events relative to the total number of congestion events. 
Each simulation is performed with $10^4$ relaxation steps and $4\times10^4$ simulation steps, 
and the autocorrelation function is calculated from the congestion sequences within the simulation steps.

\bibliography{csns20250}